\documentclass[conference]{IEEEtran}
\IEEEoverridecommandlockouts
\usepackage{cite}
\usepackage{amsmath,amssymb,amsfonts}
\usepackage{algorithmic}
\usepackage{graphicx}
\usepackage{textcomp}
\usepackage{xcolor}
\usepackage{multirow}
\usepackage{subcaption}
\usepackage{caption}
\usepackage[margin=0.7in]{geometry}
\newcommand{\RN}[1]{%
  \textup{\uppercase\expandafter{\romannumeral#1}}%
}
\usepackage[ruled,longend]{algorithm2e}

\def\BibTeX{{\rm B\kern-.05em{\sc i\kern-.025em b}\kern-.08em
    T\kern-.1667em\lower.7ex\hbox{E}\kern-.125emX}}
\begin{document}

\title{Collaborative Spectrum Sensing in Tactical Wireless Networks\\
\thanks{This research project has been supported by the NSERC-Ultra Electronics Chair on Wireless Emergency and
Tactical Communications at the \'Ecole de Technologie Sup\'erieure (\'ETS) in Montreal, QC.}
}

\author{\IEEEauthorblockN{Bryan Gingras, Ali Pourranjbar, Georges Kaddoum}
\IEEEauthorblockA{\textit{Department of Electrical Engineering} \\
\textit{\'ETS - University of Quebec, Montreal QC, Canada}\\
bryan.gingras.1@ens.etsmtl.ca, ali.pourranjbar.1@ens.etsmtl.ca, georges.kaddoum@etsmtl.ca}
}

\maketitle

\begin{abstract}
In this paper, we propose an algorithm for channel sensing, collaboration, and transmission for networks of Tactical
Communication Systems that are facing intrusions from hostile Jammers.
Members of the network begin by scanning the spectrum for Jammers, sharing this information with neighboring nodes, and
merging their respective sets of observation data into a decision vector containing the believed occupancy of each channel.
A user can then use this vector to find a vacant channel for its transmission.
We introduce the concept of nodes sharing these vectors with their peers, who then merge them into super-decision vectors,
allowing each node to better identify and select transmission channels.
We consider fading scenarios that substantially limit the reliability of the users' observations, unless they cooperate
to increase the trustworthiness of their sensing data.
We propose a pseudo-random channel selection algorithm that strikes a balance between sensing reliability with the number of
channels being sensed.
Simulation results show that the proposed system improves the network's overall knowledge of the spectrum and the
rate of Jammer-free transmissions while limiting added computational complexity to the nodes of the network, despite
the Jammers' unpredictable nature.
\end{abstract}


\section{Introduction}
The security of wireless networks is a vital concern, particularly in wireless tactical communication networks that
are subject to attacks from Jammers.
Anti-jamming techniques have been an important topic of research for many years, with jamming techniques evolving
accordingly in an endless arms race.
These attacks are particularly dangerous as they can degrade the performance of the network and possibly even cause
denial of service (DoS).
In a military context, a DoS attack can leave deployed personnel isolated and vulnerable.
Despite the constant evolution of the technologies and the ever-increasing number of anti-jamming algorithms, the core
principles behind anti-jamming remain the same, which consist primarily of avoiding Jammers or minimizing their ability
to hinder transmissions within a network.
Another approach relies on the detection of Jammers by sensing the power level at different frequencies in the
electromagnetic spectrum in order to detect signals that may be of hostile origin \cite{Grover2014}.
Ensuring that tactical wireless networks remain free of the effects of hostile jammers requires a joint optimization of
anti-jamming techniques as well as the mechanisms used by members of these networks to sense the spectrum.

Among recent research conducted in the field of anti-jamming, Slimeni \textit{et al.} \cite{Attia2018} applied Q-learning to
Secondary Users (SUs) in a Cognitive Radio Network (CRN) to teach them how long to transmit on
each channel in the spectrum before it is visited by a Jammer.
Thus, SUs do not need to continuously switch channels and can transmit on a channel for as long as the learned Jammer
schedule allows.
Zhang \textit{et al.} \cite{Zhang22018} used cooperative channel selection and power allocation to achieve multi-user and
multi-channel anti-jamming.
They accomplish this by sacrificing a fraction of some users' benefit to achieve higher overall system throughput,
effectively forcing the Jammer into a situation where its ability to jam the network is limited.

Yao \textit{et al.}\cite{Yao2019} employed decentralized collaboration and multi-agent Q-learning to select channels on which to
transmit, all while avoiding mutual interference as well as a sweeping Jammer.
They also consider that the observations of their agents are imperfect due to sensing errors.
Their solution's performance exceeds that of independent single-user Q-learning, and with a much faster convergence of
the Q-table.

Aref \textit{et al.} \cite{Aref2017} also used sweeping jamming and Q-learning in a multi-user setting to avoid mutual
interference and a sweeping Jammer, but they did not employ inter-node collaboration.
Also, unlike\cite{Yao2019}, they maintained two separate Q-tables, one for selecting channels for sensing and
another for transmission.
Jia \textit{et al.} \cite{Jia2019} considered a channel selection problem in dense wireless networks where the number of sensing
agents varies over time.
They proposed an anti-jamming dynamic game, and proved it to be an exact potential game, which guarantees the
existence of at least one pure strategy Nash equilibrium (NE).
Their approach, a ``distributed anti-jamming channel selection algorithm'', was employed, which leads to the NE of the
anti-jamming game with multiple transmitters and multiple Jammers.
This multi-agent learning algorithm runs iteratively until all sensing agents' sensing strategies (the probabilities
of sensing each channel in the next iteration of the algorithm) converge to the NE\@.

Enhancing knowledge of the spectrum occupancy can improve the performance of anti-jamming techniques.
The effect of sharing sensing information between agents is considered in \cite{Arshad2009} and \cite{Ghasemi2005}.
Arshad \textit{et al.} \cite{Arshad2009} use multi-agent collaboration in a CRN where observations are unreliable due to
fading and adverse channel conditions.
They demonstrate that multiple SUs sharing their sensing information with each other leads to significant gains in the
detection probability when compared to local sensing.
Ghasemi \textit{et al.} \cite{Ghasemi2005} address a similar topic, but they go further by exploring the impact of multi-agent
collaboration in the case of spatially correlated shadowing, where nodes who are near each other experience similar
shadowing effects.
They show that nodes that are in close proximity to each other mutually degrade their performance and lower their
probabilities of successful detection of a Primary User.

Although \cite{Attia2018,Zhang22018,Yao2019,Aref2017,Jia2019} propose solutions to the anti-jamming problem, they all
consider Jammer behaviors as fully observable, which results in them being learnable and in some cases, predictable.
Furthermore, there is clear potential in applying collaborative spectrum sensing and data fusion, such as in \cite{Arshad2009}
and \cite{Ghasemi2005}, to the particular context of tactical communications, which emphasizes anti-jamming and not only
finding available bandwidth for transmission.
Collaborative spectrum sensing, with an anti-jamming problem where Jammer behavior cannot be predicted, is a topic of
research that has not been adequately addressed.

In this paper, we propose an algorithm for channel selection, data collaboration, and transmission that leads
to improved awareness of the spectrum as well as a higher rate of unjammed transmission in a partially observable
environment where the behavior of the Jammer cannot be anticipated.
We also explore new avenues in inter-node collaboration where each node in the network shares not only its local sensing
data with its neighbors, but also its decisions with respect to the occupancy of each channel, thereby growing their
respective set of observation data that they can use to select a vacant channel on which they may safely transmit.
Finally, we introduce a collaborative channel selection algorithm that addresses unreliable sensing conditions.
Our proposed system is thus an attempt to solve the joint problem of anti-jamming and collaborative spectrum sensing.

The remainder of the paper is organized as follows.
Section \ref{model} describes the system model and the simulated environment.
Section \ref{game} provides a detailed description of the approach we take to solve the anti-jamming problem.
Section \ref{results} presents the simulation setup and results while Section \ref{conclusion} concludes the paper.

\section{System Model}\label{model}
\begin{figure}
\centering
\includegraphics[width=0.42\columnwidth]{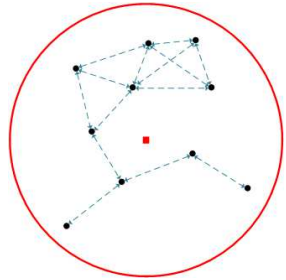}
\caption{Configuration of the network used in the simulation.
The red box in the center represents the Jammers, the red circle represents their range of operation, the black circles
represent the WNs, and the blue arrows between a pair of nodes indicate that they are neighbors and are able
to share sensing information with each other.}
\label{nw}
\end{figure}

In this work, an ad hoc network consisting of $N_{WN}$ Wireless Nodes (WNs), $N_{FB}$ orthogonal channels and a number
of Jammers equal to the number of channels is considered.
Each Jammer is assigned to one channel, and can only operate on it.
Each WN possesses a transmission range, and any WN within this range is considered its neighbor, with which it will
mutually share sensing information.
Each WN periodically senses a channel on the spectrum to observe whether or not there is a Jammer operating on that
particular channel.
It then transmits this Jammer detection observation to its neighbors, resulting in each WN in the network possessing
observations on one or more channels.
Each WN then analyzes its set of observations using a fusion rule and deduces the occupancy of each channel in the
spectrum, meaning that it will decide whether each channel is being jammed or if it is safe to use for transmission.
The WN then applies the channel selection algorithm to choose which channel it will sense next.
Finally, the WN shares its occupancy decisions with its neighbors, selects a vacant channel and uses it to broadcast data.
The WNs are dispersed geographically across a deserted terrain as displayed in Fig. \ref{nw}.

If a WN believes that all $N_{WN}$ channels are jammed, it will not attempt to transmit.
Additionally, channels for which the WN has no sensing information will be ignored by the WN, meaning that it will only
attempt to transmit on channels which it has reason to believe is vacant.
As in \cite{Lunden2013}, we assume that the actions of each WN are sufficiently synchronized to allow all of their
actions to occur simultaneously.
We further assume that all agents remain active and immobile for the entirety of the simulation, and that the number
of WNs and Jammers is fixed.
Lastly, nodes share their sensing information using a common control channel which we assume cannot be
jammed nor subject to collisions \cite{lo2012multiagent}.

Observations made by the WNs are not perfect and depend on probabilities of detection when the observed channel is
jammed, as well as probabilities of false alarm when the observed channel is vacant (meaning the WN will falsely detect
a Jammer).
Furthermore, observations are subject to signal fading.
Particularly, we separately consider Additive White Gaussian Noise (AWGN) channels as well as Rayleigh fading, both of which
differently affect the probability that a WN's observation will be incorrect.
Due to the multi-path fading component of the Rayleigh model, its impact on the sensing reliability is higher than that of
the AWGN model.
These probabilities vary with respect to the number of WNs simultaneously sensing the channel, meaning that observations
are more likely to be correct if multiple WNs sense a given channel at the same time.
When using AWGN channels, in addition to the number of users $m$ simultaneously sensing a channel, probabilities of
successful detection $p_{d,m}$ are also a function of the SNR $\gamma$ of the received signal, as shown in \eqref{eq_p_dm_awgn}.
The Rayleigh fading model considers that the WN samples the signal $N$ times, which leads to an average detection
probability $\bar{p}_{d,Ray}$ as shown in \eqref{eq_p_d_ray}.
Equations \eqref{eq_p_dm_awgn} and \eqref{eq_p_d_ray} are detailed in \cite{Digham2007}.
Rayleigh detection probabilities with $m > 1$ are given by \eqref{eq_p_dm_ray}.
In all cases, $\sigma^2$ is the variance of the sampled signal, $a$ is a non centrality parameter, and $\lambda$ is a decision
threshold.

\begin{equation}
p_{d,m,AWGN} = Q_{mN/2} \left(\sqrt{\frac{a\gamma}{\sigma^2}}, \sqrt{\frac{\lambda}{\sigma^2}}\right),
\label{eq_p_dm_awgn}
\end{equation}

\begin{equation}
\begin{split}
\bar{p}_{d,Ray} = e^{-\frac{\lambda}{2\sigma^2}}\sum_{i=0}^{N/2-2}\frac{\left( \frac{\lambda}{2\sigma^2} \right)^i}{i!}+\left( \frac{2\sigma^2+a\bar{\gamma}}{a\bar{\gamma}} \right)^{N/2-1}\\ \times
\left(e^{-\frac{\lambda}{2\sigma^2+a\bar{\gamma}}} - e^{-\frac{\lambda}{2\sigma^2}} \sum_{i=0}^{N/2-2}\frac{\left( \frac{\lambda a \bar{\gamma}}{2\sigma^2(2\sigma^2 + a\bar{\gamma})} \right)^i}{i!}\right),
\label{eq_p_d_ray}
\end{split}
\end{equation}

\begin{equation}
\bar{p}_{d,m,Ray} = 1 - \prod_{i=1}^m (1 - \bar{p}_{d,Ray,i}).
\label{eq_p_dm_ray}
\end{equation}

The power of the Jammer signal received by the user is given by $P_T\times(d/d_0)^{\phi}$,
where $d/d_0$ is the distance between the transmitter and the receiver divided by a reference distance, the constant
$\phi$ corresponds to an attenuation factor which depends on the physical environment in which the transmission takes
place, and $P_T$ represents the initial power of the transmitted signal.
We consider this received power to include noise as well as the channel gain in the case of Rayleigh fading.
WNs must therefore cooperate in order to increase the diversity order $m$ and obtain more reliable sensing information.

\begin{figure}
\centering
\includegraphics[width=0.70\columnwidth]{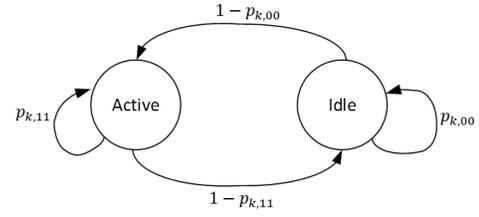}
\caption{Two-State Jammer Markov Model.
Upon initialization, a Jammer's initial state is decided randomly.
At each time step, it may remain in its current state according to probabilities $p_{k,00}$ and $p_{k,11}$ or
switch to the alternate state.}
\label{markov}
\end{figure}

At any given moment, a Jammer may be idle, or it may be actively jamming the channel in order to intercept transmissions
between WNs.
We represent its behavior using a Markov model with states 0 and 1, which respectively signify idle and active.
We define $p_{k,00}$ as the probability that, at each time step, Jammer $k$ remains in the idle state, and $p_{k,11}$ as
the probability that it stays in the active state.
It follows that $p_{k,01} = (1 - p_{k,00})$ is the probability that the Jammer goes from idle to active, and
that $p_{k,10} = (1 - p_{k,11})$ is the probability of going from active to idle.
The corresponding Markov model is represented by Fig. \ref{markov}.

\section{Stochastic Game Formulation}\label{game}
Each WN can only sense one channel at a time, therefore making it impossible for it to observe the entire spectrum at once.
The multi-agent jamming problem can be represented as a partially observable stochastic game modelled
as follows:
\begin{itemize}
    \item A set of time steps $t = 1, 2, \ldots, T$ that are each of fixed length and split into three sub-slots:
    sensing, collaboration, and transmission.
    \item A group of Wireless Nodes $M=\{1,\ldots,N_{WN}\}$ dispersed throughout the environment.
    \item A group of $N_{FB}$ channels in the frequency band.
    Each WN senses exactly one of these channels at each time step $t$ and observes its occupancy.
    \item A set of possible occupancy values for a channel $j$ as it would be sensed by WN $i$ at time step $t$,
    given by $s^{i,j}_t \in \{vacant, occupied\}$.
    \item The vector $\pmb{s}^i_t$ represents the occupancy of each channel as it would be seen by WN $i$ at
    time step $t$, given by $\pmb{s}^i_t=[s^{i,1}_t, \ldots, s^{i,N_{FB}}_t]$.
    Its value is a member of the set $\{vacant, occupied\}$.
    \item The state $\pmb{s}_t$ is an array of vectors $\pmb{s}^i_t$, showing each WN's perception of each channel
    at time $t$, given by $\pmb{s}_t = [\pmb{s}^1_t, \ldots, \pmb{s}^{N_{WN}}_t]$.
    \item A set of possible sensing actions $a^i_t \in A_i$, where $i \in M$, and $A_i=\{1,\ldots,N_{FB}\}$ represents
    which channel will be sensed by the WN at time step $t$.
    \item $\tau^i_t$ represents an observation made by WN $i$ at time $t$, i.e.\ it is the observation made from
    taking action $a^i_t$.
    Its value is a member of the set $\{vacant, occupied\}$.
    Due to the imperfect nature of the WNs' observations, this value may not reflect the actual occupancy of the
    channel as given by $s^{i,j}_t$.
    \item A vector of sensing decisions made by each WN $i$ $\pmb{d}^i_t \in \pmb{D}_{i,t}$.
    Each element of $\pmb{d}^i_t$ is a member of the set $\{vacant, occupied\}$.
    \item A vector of super-decision vectors $\pmb{\delta}^i_t$ obtained by combining WN $i$'s
    decision vector $\pmb{d}^i_t$ with its neighbors' decision vectors $\pmb{d}^j_t \ \forall \ j$ into a single vector.
    Like $\pmb{d}^i_t$, each element of $\pmb{\delta}^i_t$ is a member of the set $\{vacant, occupied\}$.
    \item A transmission outcome $x^i_t \in \{successful, jammed\}$ representing the outcome of the transmission
    performed by WN $i$ at time step $t$.
    \item A state transition function $\phi$: $\pmb{S} \times \pmb{A} \times \pmb{S}\rightarrow\mathbb{R}$ that
    defines the state transition probabilities $P(\pmb{s}_{t+1}\ |\ \pmb{s}_t,a^1_t, \ldots, a^{N_{WN}}_{t})$.
    We use the Jammer probabilities $p_{k,00}$ and $p_{k,11}$ as the state transition probabilities.
    We assume that the agents' actions do not affect the state transition probabilities:
    $P(\pmb{s}_{t+1}\ |\ \pmb{s}_t,a^1_t, \ldots, a^{N_{WN}}_{t}) = P(\pmb{s}_{t+1}\ |\ \pmb{s}_t)$.
    In other words, we assume that an WN's sensing action does not cause any interference on the channel that could
    affect the observation of another WN.
\end{itemize}

\vspace{-0.1cm}

\subsection{Spectrum Sensing, Cooperation and Fusion}
A time step of the simulation can be summarized by the following.
Each WN $i$ senses a channel and receives a sensing observation datum $\tau^i_t$, which it then shares with
each of its neighboring WNs $j$.
Each WN $i$ then combines its observations into a decision vector $\pmb{d}^i_t$ by applying a fusion rule on its set of
observation data.
Then, basing itself on the outcome of the observation that it has made, as well as the actions taken in the previous
time step by its neighbors, each WN chooses which channel to sense in the following time step $t + 1$.
Next, in order to improve their knowledge of the presence of Jammers in the spectrum, the WNs transmit their
respective decision vectors $\pmb{d}^i_t$ to their neighbors, who again apply a fusion rule, resulting in a
super-decision vector $\pmb{\delta}^i_t$.
In doing so, an WN $i$ not only synthesizes the observations made by itself and its neighbors $j$, but also gains
indirect access to the observations made by $j$'s neighbors that are outside $i$'s transmission range, thereby
increasing the ranged of shared information from one hop to two.
Finally, each node uses $\pmb{\delta}^i_t$ to select a channel it believes to be vacant and transmits on it.
This is summarized in Algorithm 1.
\begin{algorithm}
\caption{Sensing, Cooperation and Fusion Algorithm}
    \begin{algorithmic}
        \STATE Initialize $t = 0$
        \STATE Initialize set of Wireless Nodes $M$
        \STATE Initialize set of Jammers $J$
        \STATE Initialize set of Channels $W$
        \STATE Initialize $a^i_0 = random(1, \ldots, N_{FB}), \ i \in \{1, \ldots, N_{WN}\}$
        \STATE Initialize $w^z_0 = bernoulli(0,5), \ z \in \{1, \ldots, N_{FB}\}$
        \STATE $s_0 = computeOccupancy(M, J, W)$
        \WHILE {$t < T$}
            \STATE $\tau^i_t = sense(a^i_t),\ i \in \{1, \ldots, N_{WN}\}$
            \FOR {each WN $i$ in $M$}
                \STATE Transmit tuple $\{a^i_t, \tau^i_t\}$ to neighbors
                \FOR {each neighbor $j$ of WN $i$}
                    \STATE $j$ receives $\{a^i_t, \tau^i_t\}$ from $i$
                    \STATE $j$ adds $a^i_{t+1}$ to its action vector: $\pmb{a}^j_{t+1} = \pmb{a}^j_{t+1} \cup a^i_{t+1}$
                \ENDFOR
            \ENDFOR
            \STATE $\pmb{d}^i_t = fusion(\tau^i_t, \{\tau^j_t\}_{\forall j}), i \in \{1, \ldots, N_{WN}\}$
            \STATE $a^i_{t+1} = chooseAction(\tau^i_t, \pmb{a}^i_t), i \in \{1, \ldots, N_{WN}\}$
            \FOR {each WN $i$ in $M$}
                \STATE Transmit decision vector $\pmb{d}^i_t$ to neighbors
                \FOR {each neighbor $j$ of WN $i$}
                    \STATE $j$ receives $\pmb{d}^i_t$ from $i$
                \ENDFOR
                \STATE $\pmb{\delta}^i_t = fusion(\pmb{d}^i_t, \{\pmb{d}^j_t\}_{\forall j})$
                \STATE $x^i_t = transmit(\pmb{\delta}^i_t)$
            \ENDFOR
            \STATE $s_{t+1} = computeOccupancy(M, J, W)$
            \STATE $t = t + 1$
        \ENDWHILE
    \end{algorithmic}
\end{algorithm}

We repurpose the transmission sub-time slot selection algorithm from \cite{Lunden2015} for our sensing channel selection
algorithm.
The algorithm functions as follows: if a WN detects a Jammer on a
given channel, it will sense that channel again during the following time slot $t + 1$.
If the sensing action does not detect a Jammer, then we enter an exploration-exploitation scenario, where there is a
probability $\epsilon_n \in [0,1]$ that the WN will exploit its neighbors' knowledge of the spectrum by selecting the
action $a^j_{t}$ of one of its neighbors $j$.
If taking action $a^j_{t}$ yields a Jammer, then $j$ is certain to choose this action again for the next time slot
$t + 1$, and if WN $i$ chooses this action as well for the same time slot, then $i$ and $j$ assist each other by
increasing the diversity order $m$ and therefore the detection probability $p_{d,m}$ of both WNs.
If the observation $\tau^j_t$ is a false positive, then a repeated observation on that channel with a higher value of $m$
(and therefore a lower probability of false alarm) will be more likely to yield the correct observation of $vacant$.

If the two previous criteria are not met, the third possibility is for the WN to select a channel that was not sensed by
itself or by any of its neighbors, in order to explore a greater range of channels that are not currently being sensed
by the WN or its surrounding nodes.
The value of $\epsilon_n$ is therefore a trade-off between sensing reliability and the number of channels being sensed
at any given moment.
This is summarized in Algorithm 2.

\begin{algorithm}
\caption{Pseudo-Random Algorithm for Action Selection}
    \begin{algorithmic}
    \IF {$\tau^i_t == occupied$}
        \STATE $a^i_{t+1} = a^i_t$
    \ELSIF {$u(0,1) \leq \epsilon_n$}
        \STATE $a^i_{t+1} = a^j_t$
    \ELSE
        \STATE $a^i_{t+1} = random(A_i \ \backslash \ \{a^i_t,\ a^j_t \ \forall \ j\})$
    \ENDIF
    \end{algorithmic}
\end{algorithm}

The fusion strategy used by the WNs to merge their sensing information into a decision vector $\pmb{d}^i_t$ is the
OR rule, meaning that each WN considers a channel to be jammed if its own observation,
or one that it receives from at least one of its neighbors, indicates that the channel is jammed.
The WNs apply this same OR fusion rule when combining their decision vectors into a super-decision vector
$\pmb{\delta}^i_t$.

\section{System Performance}\label{results}
Since the major contribution of this article lies in the algorithm used for action selection, we judge the efficacy of
our solution by comparing the performance of this algorithm against a collaborative multi-agent reinforcement learning
scheme that attempts to build a channel selection policy based on sensing data as proposed in \cite{Lunden2013}.
Additionally, we compare our results with those obtained using randomly chosen actions.

We evaluated the performance of the algorithms described above using a simulation of a deserted, empty terrain containing
a set number of WNs and Jammers implemented in Python 3.5.2.
The simulation ran for a pre-set number of time steps $T$ according to the stochastic game described above.
Throughout the simulation, we computed the number of detected incidences of jamming over the total number of times
that jamming has occurred up until the current time step.
As a secondary performance metric, we also calculated the number of successful transmissions over the total number of
transmissions that had been attempted up to and including the current time step.
Since each WN attempts to transmit once per time step, we knew this latter value ahead of time to be simply
$N_{WN} \times t$.

The scenario used for evaluation consisted of 10 WNs and 10 Jammers (and therefore 10 channels).
For each Jammer, the values of $p_{k,00}$ and $p_{k,11}$ were randomly generated between bounds of $0.85$ and $0.98$ as in \cite{Lunden2013}.
Due to the high values of these probabilities, a Jammer $k$ was much more likely to remain in its current state from one
time step to the next than to transition to the alternate state.
The simulation was run for $T = 2000$ time steps.
Transmitted signals were attenuated at a rate of $P_T\times(d/0.05)^{-2.3}$.
We used the exponent -2.3 to represent a flat, empty environment much like a desert or field.
The WNs were arranged across the terrain as per Fig.~\ref{nw}.
We also used a reference distance of $0.05 \ km$ \cite{Lunden2013}.
We used the following false alarm probabilities: $p_{fa,1}=0.0015$, and $p_{fa,n}=10\textsuperscript{-7}$
for $n \geq 2$ for AWGN channels, as well as $p_{fa,1}=0.83$, $p_{fa,2}=0.32$, $p_{fa,3}=0.03$, $p_{fa,4}=0.003$,
and $p_{fa,n}=0.001$ for $n \geq 5$ \cite{Lunden2013} for Rayleigh fading channels.

\begin{figure}
\centering
    \begin{subfigure}[b]{0.45\textwidth}
        \includegraphics[width=0.85\textwidth]{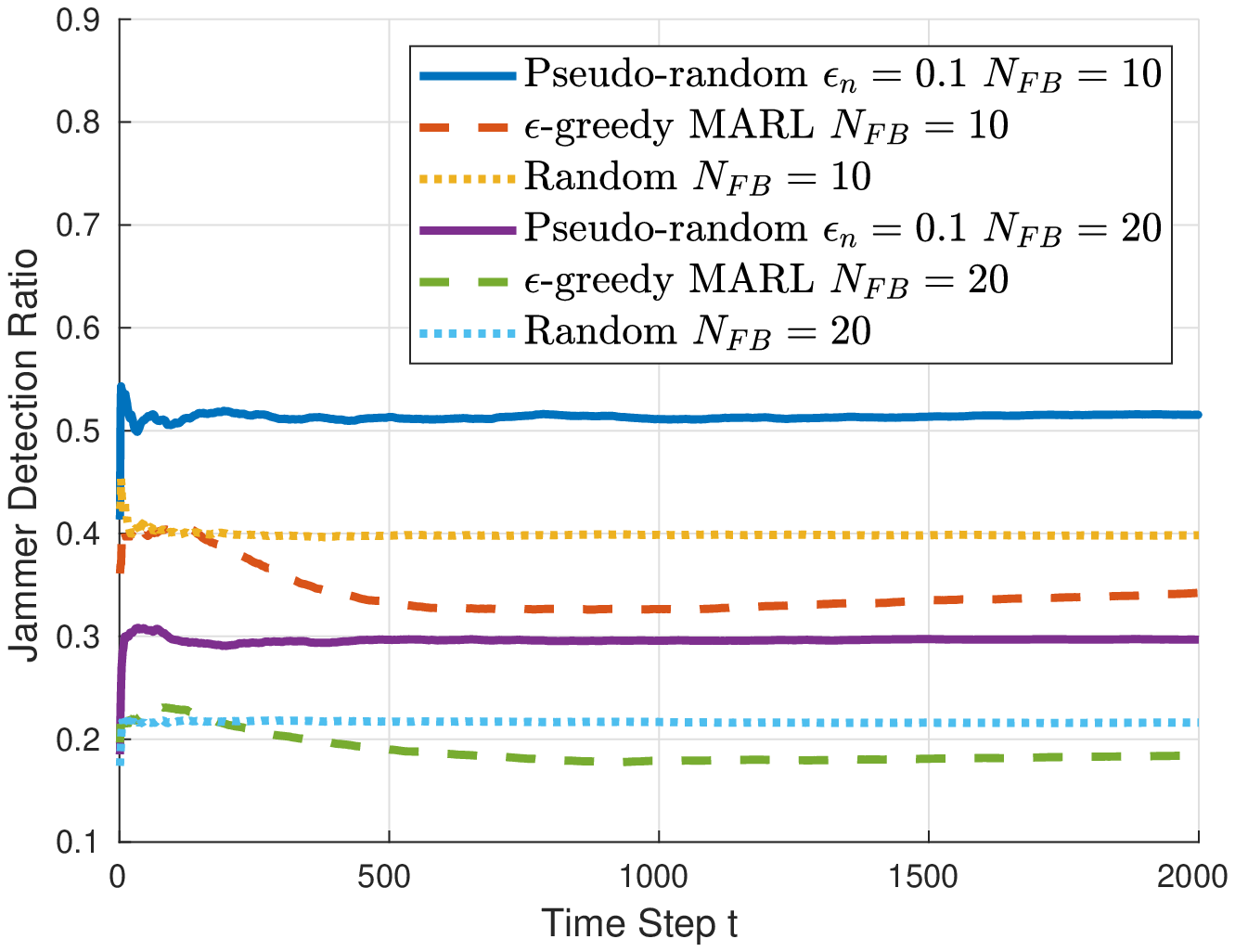}
        \caption{}
        \label{awgn}
    \end{subfigure}
    \begin{subfigure}[b]{0.45\textwidth}
        \includegraphics[width=0.85\textwidth]{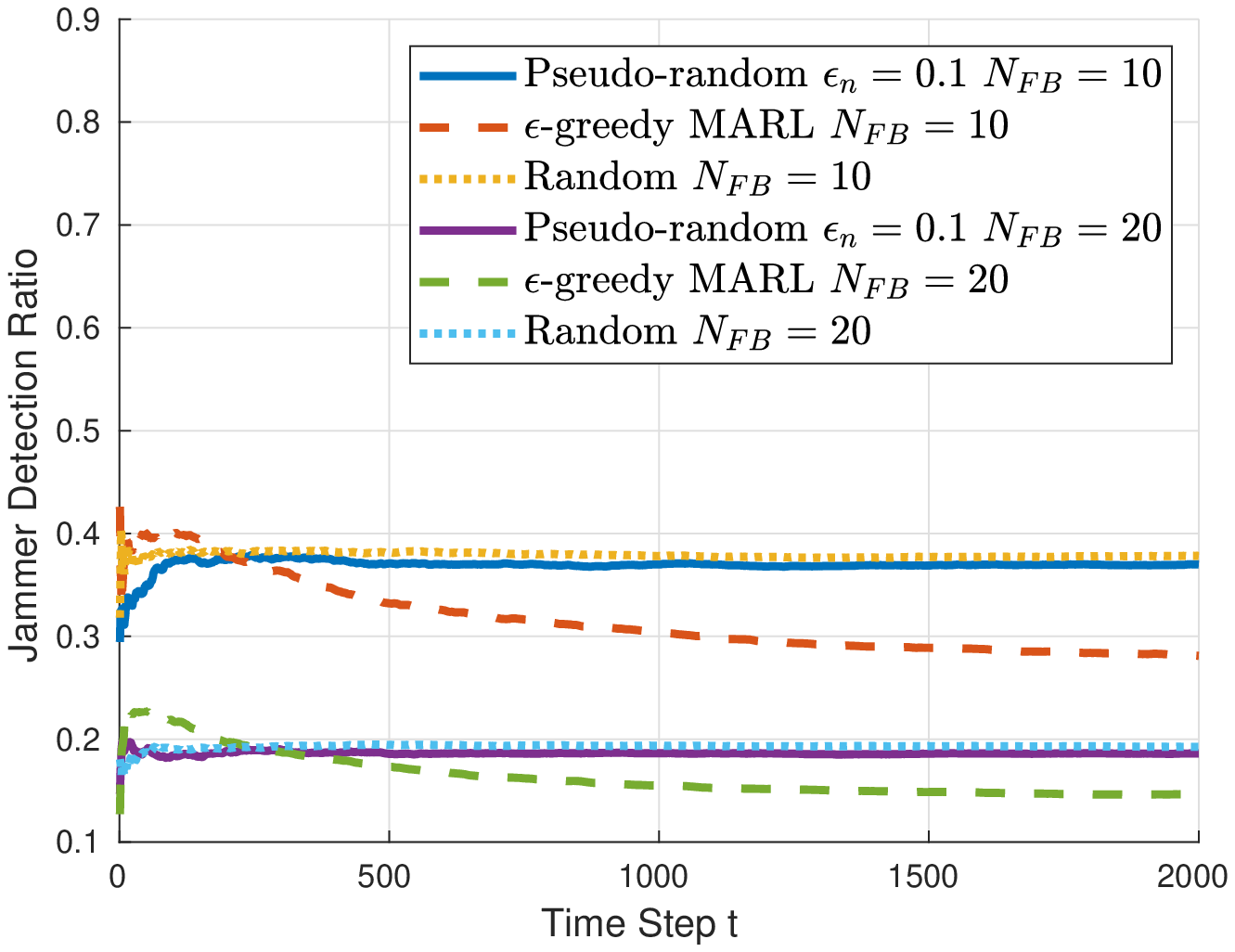}
        \caption{}
        \label{ray}
    \end{subfigure}
    \caption{Performance evaluation using Jammer Detection Ratio with (a) AWGN channels and (b) Rayleigh fading.}
\end{figure}

The Jammers' transmission power was held constant at 15 dB\@.
Using values of received SNR by the WNs ranging from 0 dB to 15 dB and diversity orders ranging from 1 to 6 WNs simultaneously sensing the
same channel, we used \eqref{eq_p_dm_awgn} to compute a 2D matrix that served as a look-up table for the different values of
$p_{d,m,AWGN}$ that the simulation software could consult every time a WN sensed a channel.
Similarly, we used \eqref{eq_p_d_ray} to compute a 1D array giving detection probabilities $\bar{p}_{d,Ray}$ where $m = 1$,
using the same range of SNR values and when using the Rayleigh model for channel fading.
Detection probabilities with Rayleigh fading when $m > 1$ were dynamically calculated during the simulation using \eqref{eq_p_dm_ray}.
The variance $\sigma^2$ is equal to $1$, $a$ is set to $2$, and $\lambda$ is equal to $12.1$ in order to approximate
probabilities seen in \cite{Lunden2013}.

Since we used an OR fusion rule, a single false alarm could lead to an incorrect decision when using this strategy, but
the false alarm probabilities used in this simulation were generally smaller than the values of $p_{d,m,AWGN}$ and $\bar{p}_{d,m,Ray}$
that we generally observed, which means that most observations where $\tau^i_t$ was equal to $occupied$ were due to correct
observations and not false alarms, though it is important to note that observations are less reliable when using Rayleigh
fading and were thus more prone to incorrect decisions than when using AWGN channels.

In Figs. \ref{awgn} and \ref{ray} we illustrate the performance of our algorithm with $\epsilon_n = 0.1$ as well as the
multi-agent reinforcement learning scheme and random action selection using AWGN channels and Rayleigh fading, respectively.
The probabilities of a false alarm were higher for Rayleigh fading, which explains why the algorithms generally show
lower performance in Fig. \ref{ray} than in Fig. \ref{awgn}.

\begin{figure}
\centering
    \begin{subfigure}[b]{0.45\textwidth}
        \includegraphics[width=0.85\columnwidth]{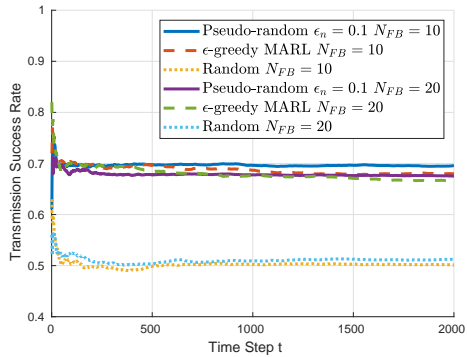}
        \caption{}
        \label{local}
    \end{subfigure}
    \begin{subfigure}[b]{0.45\textwidth}
        \includegraphics[width=0.85\columnwidth]{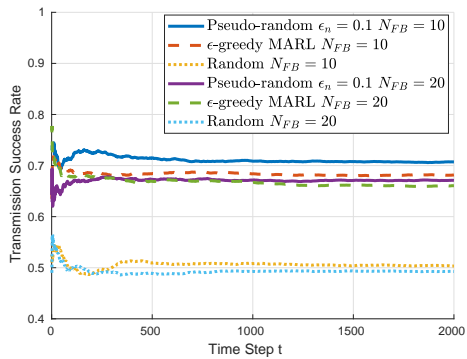}
        \caption{}
        \label{super}
    \end{subfigure}
    \caption{Performance evaluation using Transmission Success Rate with (a) Local Decision Vectors and (b) Super-Decision Vectors.}
\end{figure}

Next, in Figs. \ref{local} and \ref{super}, we observe the rate of successful transmission, i.e.\ the number of
transmissions that took place on a vacant channel over the total number of attempted transmissions $N_{WN} \times t$,
using simple decision vectors $\pmb{d}^i_t$ versus super-decision vectors $\pmb{\delta}^i_t$.
Furthermore, all of the simulations used to determine the transmission success rate were done using AWGN channels.
Time steps where the WN could not find a vacant channel for transmission were excluded from the calculation.
Each curve in each figure is an average of 100 iterations of the simulation.

\vspace{-0.1cm}

\subsection{Jammer Detection}\label{jdr}
We remark in Fig. \ref{awgn} that the pseudo-random algorithm out-performed the multi-agent reinforcement
learning scheme and random action selection.
The algorithm exploited the fact that jammed channels were likely to remain jammed, and would continue to sense the channel
until it suddenly became vacant.
In such an event, the WN could, according to $\epsilon_n$, assist its neighbor or sense an unsensed channel in
the hopes of finding a new channel that it could exploit over several time steps.
In \ref{ray}, the performance of the pseudo-random algorithm is near that of the random action selection algorithm.
The higher likelihood of incorrect observations when using Rayleigh can give rise to a situation where the WN suddenly
senses a jammed channel as being vacant, which causes it to stop sensing that channel and sense a different, potentially
vacant channel.

Due to the random nature of the Markov model controlling the Jammers' state transitions, it is impossible to anticipate
when a Jammer will become idle or active.
The inherent randomness of the Jammer may explain how random action selection performs as well as it does.
This also explains the MARL algorithm's difficulty to devise a useful transmission policy.
When we increased $N_{FB}$ from $10$ to $20$, we essentially doubled the number of possible instances of jamming that could
occur during the simulation, and considering that the number of WNs remains the same, we observe the expected result which
is for the performance of each algorithm to drop by half when doubling $N_{FB}$.

\subsection{Transmission Success Rate}\label{tsr}
In Fig. \ref{local}, we can observe that the relative performance of each algorithm is fairly similar to that of
Fig. \ref{super} in terms of the rate of successful transmission.
We also notice that the performance of an algorithm does not vary considerably when $N_{FB} = 10$ versus $N_{FB} = 20$.
This can be attributed to the fact that a WN only transmits on channels for which it possesses information on its
occupancy.
In other words, even if the number of channels for which the WN has no information increases, the number of sensing
data that it possesses does not change due to the fact that $N_{WN}$ remains constant at $10$.
However, the average diversity order of each sensing action may be reduced if $N_{FB}$ is raised to $20$, resulting in
less reliable observations, which may explain why the cases where $N_{FB} = 10$ generally perform slightly better than
when $N_{FB} = 20$.
Therefore, increasing $N_{FB}$ beyond $N_{WN}$ does not vastly affect the transmission success rate.

We note that in both figures, the performance roughly does not exceed 70\%, which may be an asymptotic value that
is a function of $N_{WN}$ as well as other factors including the values of the probabilities of false alarm.
Despite this, we see that using super-decision vectors $\pmb{\delta}^i_t$ performs slightly better than when using
local decision vectors.
However, this simulation does not consider the added overheads, both in terms of time and bandwidth, that arises when
every WN shares its local decision vector $\pmb{d}^i_t$ with its neighbors.
Particularly, it is important to note that extending the collaboration phase of each time step reduces the
amount of time available for the WNs to transmit data, which would be a crucial consideration if this
simulation modeled time as being continuous instead of a succession of time slots.
Lastly, it is interesting to note that despite the fact that the system model is geared towards improving the Jammer
detection ratio, our algorithm still leads to an improved transmission success rate, which gives
credence to the idea that improved Jammer detection translates into safer transmissions.

\vspace{-0.11cm}

\section{Conclusion}\label{conclusion}
In this article, we demonstrated the effectiveness of a simple, yet promising, pseudo-random channel sensing algorithm
for tactical wireless networks along with an innovative data collaboration scheme that makes better use of neighboring nodes'
sensing information using super-decision vectors.
Awareness of spectrum usage favors Jammer-free transmissions between Wireless Nodes, which is an important
consideration for ensuring the safety and effectiveness of military personnel on the field.
When coupled with appropriate data fusion algorithms, this algorithm leads to a higher performance with respect to the
rate of Jammer detection and the number of non intercepted transmissions compared to random channel selection as well as
multi-agent reinforcement learning.

\vspace{-0.1cm}

\bibliographystyle{IEEEtran}
\bibliography{css}

\end{document}